\documentclass{ws-book9x6}

\usepackage{ws-book-thm}   
\usepackage[pdfpagelabels=false,colorlinks=true,allcolors=black]{hyperref}  

\usepackage{footmisc} 

\usepackage[numbers]{natbib}

\title{Artificial Intelligence for Science}      

\begin{document}










\setcounter{page}{1}

\setcounter{chapter}{6}
\chapter[Machine Learning for Complex Instrument Design and Optimization]{Machine Learning for Complex Instrument Design and Optimization \\ 
{\small Barry C. Barish, Jonathan Richardson, Evangelos E. Papalexakis,}\\
\small{Rutuja Gurav (University of California, Riverside)}}
\label{ch1}

\begin{figure}[h!]
    \centering
    \includegraphics[width=1\textwidth]{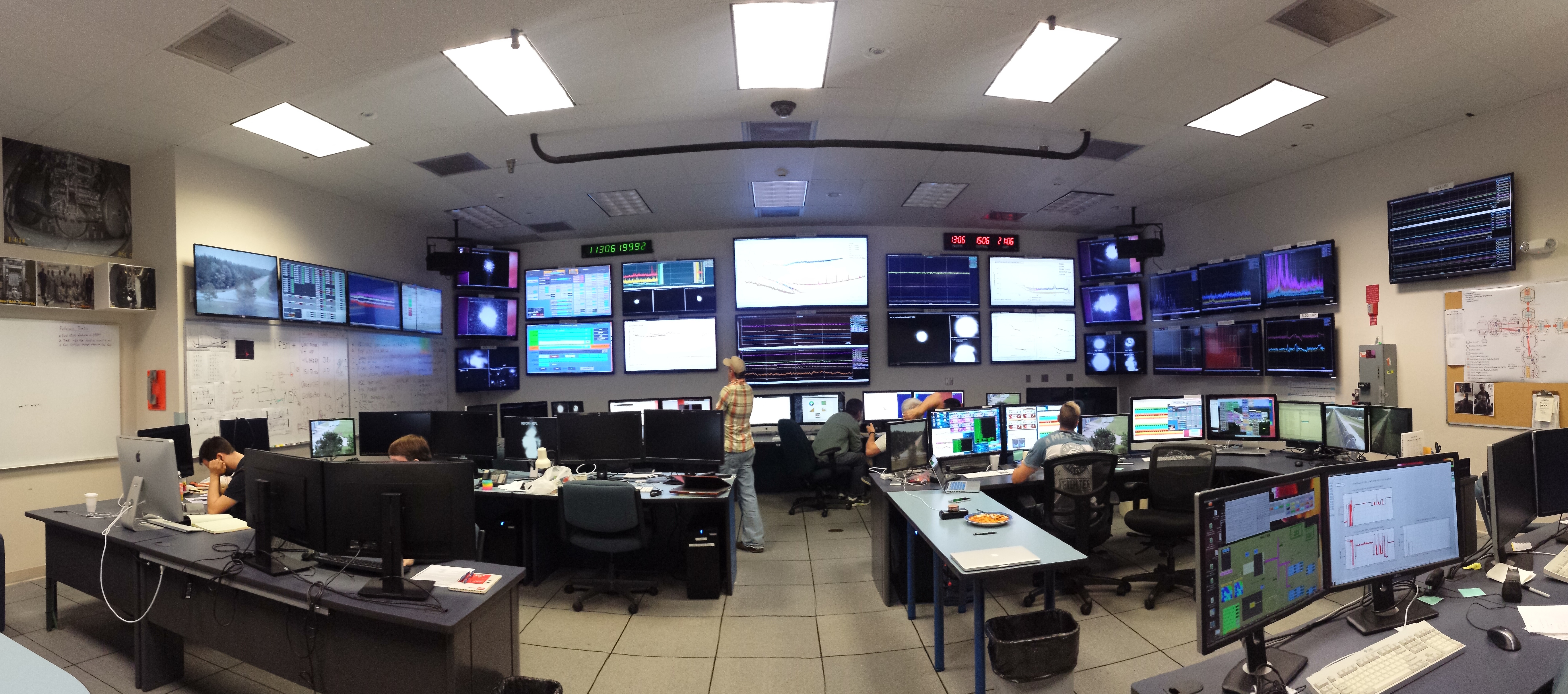}
    \caption{Inside the control room of the Laser Interferometer Gravitational Observatory (LIGO) in Livingston, Louisiana. Image credit: Amber Stuver/Wikimedia Commons.}
\end{figure}

\section{Abstract}
In modern experimental physics, particle accelerators and gravitational-wave observatories enable a wide-range of research at the frontiers of science. These instruments are highly complex consisting of many interacting systems which can face significant operational challenges. Apart from the experiment's main data product, a lot of data about the experimental apparatus and its environment is recorded. Machine learning techniques can analyze this big data at scale and find useful insights into operational faults potentially improving the instrument's performance and achieving the design goals. Speaking of design, machine learning can also accelerate/augment the expensive physics simulations used during the design phase of such large-scale instruments.
\section{Introduction}
\label{intro}
In the era of large-scale scientific experiments, big data management and high performance computing have become indispensible for the fast storage, retrieval, and analysis of the vast amounts of data generated. At the same time, there has been a growing interest in using advanced machine learning (ML) techniques for data analysis to make scientific discoveries. However, the potential of ML to accelerate scientific discovery is not limited to the analysis of an experiment's main data products. Frontier experimental apparatuses like the Large Hadron Collider (LHC), Laser Interferometer Gravitational-Wave Observatory (LIGO) and Electron-Ion Collider (EIC) are highly complex instruments with hundreds of degrees of freedom stabilized by cross-coupled feedback servos and with thousands of auxiliary sensors. The 4~km long detectors of LIGO, for example, consist of six coupled laser cavities formed by dozens of mirrors suspended from quadruple-stage pendula and mounted on active seismic isolation platforms. LIGO's main data product is the \textit{strain} \footnote{\textit{Strain} is the fractional space change across a 4~km long arm of the interferometer relative to the total length of the arm.} which is used by astrophysicists to search for gravitational-wave signals. In addition to the main strain channel, each LIGO detector records over 10,000~channels monitoring the operation of each subsystem and the seismic, acoustic, and electromagnetic environment (for an overview of LIGO's environmental monitoring system, see \cite{Nguyen:2021}). The complexity of large scientific instruments, with their vast quantities of auxiliary data, presents an emerging opportunity to use ML more broadly to learn about the instrument itself.

\begin{figure}[t]
    \centering
    \includegraphics[width=1\textwidth,trim={2mm 2mm 2mm 0},clip]{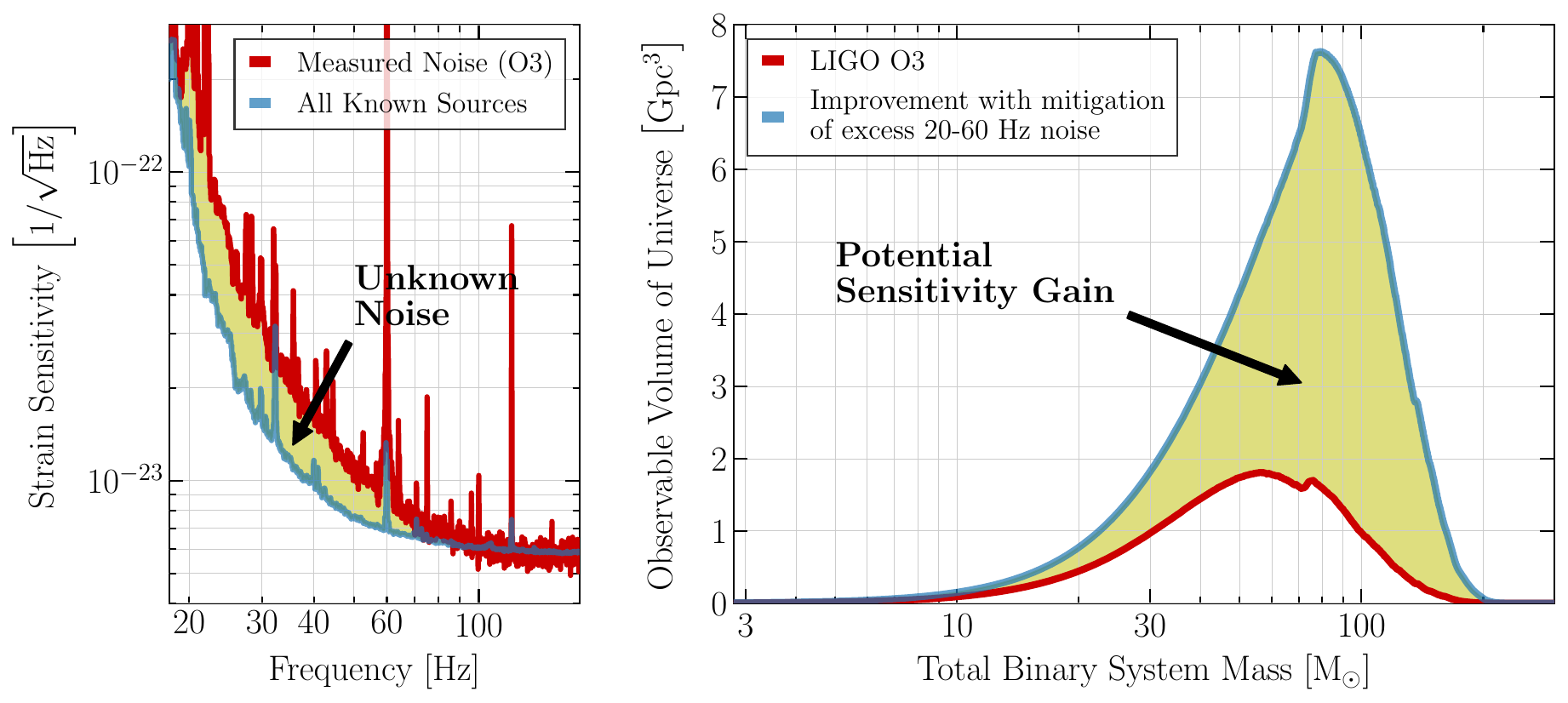}
    \caption{\textit{Left:} Noise floor of the LIGO Hanford gravitational-wave detector during the O3 observing run (red), compared to the budgeted detector noise (blue). As shown, most of the noise in the 20--60~Hz band remains unidentified. \textit{Right:} Projected improvement in the observable volume of Universe for equal-mass binary black hole mergers, as a function of the total mass (in the source frame) of the binary, if this excess noise were identified and fully mitigated.}
    \label{fig:noise}
\end{figure}

Leveraging ML tools to diagnose performance limitations, such as poorly understood instrumental noise or control instabilities, and to identify more optimal designs could lead to big scientific returns through improved sensitivity, operational up-time, and data quality. As one example, nonlinear, or nonstationary, noise couplings of mostly unknown origin now limit the Advanced LIGO detectors in several ways. In particular, most of the detector noise in a  band of key scientific interest, 20–60 Hz, remains unidentified altogether. The unidentified excess noise is shown in the left panel of Figure~\ref{fig:noise}, indicated by the shaded region. The right panel of Figure~\ref{fig:noise} show the impact of this noise on a key astrophysical metric. As shown, identifying and mitigating this noise would significantly enlarge both the volume of universe and the astrophysical mass range accessible to gravitational-wave science, including enabling observations of mergers of the most massive stellar-origin black holes.

The unidentified excess noise in the LIGO detectors is believed to originate through nonlinear mechanisms because it does not exhibit high coherence with any of the multitude of auxiliary signals. For this reason, it has proven very difficult to pinpoint the origins of nonlinear noise in Advanced LIGO using traditional (linear time-invariant) system identification techniques. However, such challenges are far from being specific to LIGO. In fact, they are common to complex actively controlled instruments, with another example being particle accelerators. In accelerators, ML can be used to optimize the performance of colliding beams, reduce beam losses, and steer the beam.

Developing new tools and approaches for understanding instrumental noise, as well as other operational anomalies, is thus critical to the future of complex instruments such as LIGO, the LHC, and the EIC. New unsupervised ML methods to model and analyze the multitude of auxiliary signals recorded in the detectors, over extended periods of time, offer an emerging opportunity to generate {\it actionable~insights} that will guide instrument commissioning efforts and future design decisions. Previous applications of ML in complex systems like LIGO~(for a through review, see~\cite{Cuoco:2020}) or accelerators~\cite{edelen2018opportunities}, overwhelming employ a supervised learning paradigm, where the desired input-output behavior of a system is known and is being approximated by a ML model (e.g., estimating the masses of the two colliding black holes in an observed gravitational wave). In many of the problems encountered within complex instrument operations, such supervision is scarce or non-existent, calling for the development of novel ML methods to tackle them. Similar challenges also arise in the operation of large complex systems in general~\cite{khan2018review}, so advances in this area will open opportunities for unsupervised and explainable knowledge discovery at-large.

In the remainder of this chapter, we first discuss the key challenges and requirements for developing unsupervised learning models for complex instruments. This section also introduces ML terminology and concepts on which the later sections of this chapter will rely. We then discuss the potential of ML as a tool for diagnosing and optimizing the operational performance of an experimental apparatus, using the LIGO detectors as an illustrative case study. Finally, we discuss the prospect of using ML for instrument design, with applications to optimizing existing accelerators like the LHC as well as the ground-up design of future frontier accelerators such as the EIC\footnote{Find more information about the EIC here - \url{https://www.bnl.gov/eic/}}.
\section{Machine Learning Challenges and Requirements}
\label{MLConcepts}

In adapting or developing new ML methods for complex instrument operations management and design, there are a number of fundamental challenges that must be addressed. Below we provide a brief overview of those challenges. Although they are presented in a serial and independent manner, typically a combination of them manifests in any problem we may encounter.

\paragraph{Weak or limited supervision}
\label{ml-limsup}
Supervision is a key aspect of machine learning, and the paradigm of \emph{supervised learning} is, most likely, the one to which most readers would have already been exposed: there is an abundance of (input data, desired output) pairs (e.g., an image and its associated label), and the task is to learn a model that maps a representation of the input to the desired label in a manner that {\em generalizes} to unseen inputs. However, a vast number of problems in the context of instrument diagnostics and design are not afforded the amount of human annotations necessary to train such models. For instance, when predicting operational failures, the annotations that would be most desirable would be ones which would characterize the parts of the instrument which are responsible for the failure. Unfortunately, this presents a {\it Catch-22}, since those annotations are the knowledge which we seek to extract from our application of machine learning. In such a case, we may resort to \textit{proxy data} which can be obtained cheaply, do not require extensive human annotation, and may help us train models which can shed some light on the ultimate intractable task. We may use the presence or absence of instrument failure as a coarse label that indicates normal or abnormal operation, in hopes that the model that is able to successfully predict such a coarse label is capturing useful information that can enable exploratory analysis to further characterize such a failure. Finally, this challenge is compounded by the fact that, even though the amounts of data generated by an instrument are massive, typically when creating ML-friendly datasets from which a model can learn meaningful and generalizable patterns there is significant cleaning, down-selecting of channels and time periods, and other forms of pre-processing involved, which result in substantially reduced sizes for available curated datasets. This makes learning more challenging.

\paragraph{Explainable models}
\label{ml-exp}
Machine learning models which are tasked with mapping input feature representations (e.g., sensor readings that monitor the state of the instrument for a certain time window) to a desired prediction (e.g.,  failure at time $t_{\rm f}$) can take many different forms, all the way from classification trees and parametric equations to deep neural networks. As we mentioned above, a fundamental requirement for any such model is to generalize well to unseen examples. However, this is only one of the dimensions in which one can examine machine learning models. A dimension that is vital to many problems encountered in the context of this chapter is \emph{explainability}, which measures the degree to which a human can readily understand the reasoning behind the model’s decision and attribute such decision to different parts of the input data (in our running example, such attribution could be highlighting different sensors and temporal windows as bearing most of the weight of the prediction of a failure). Simple linear models or tree-based models lend themselves directly to such explanations (a linear model assigns different weights to inputs, and a tree can provide a set of rules based on the inputs which led to the decision). As machine learning models move towards highly-overparametrized deep neural networks, the ability to generalize successfully increases. However, the ability to provide such explanations and attributions to the input becomes increasingly more challenging. Thus, there is an active research area within machine learning which is concerned with different paradigms and methods of explainability, ranging from readily explainable models to models for which only post-hoc explanations can be provided \cite{guidotti2018survey}. In our context, any such form of explanation can be crucially important: coupled with the challenge of weak supervision described above, imbuing explainability to a model tasked with predicting an auxiliary target (e.g., whether a failure occurred) can shine light on parts of the data which can then be further examined systematically by domain experts. In this way, we may gain insight into potential physical mechanisms that can be responsible for the observed outcome (e.g., instrument failure).

\paragraph{Theory/physics-guided models}
\label{ml-physguided}
A major promise of modern deep learning models is their purported ability to capture arbitrarily non-linear functions between input and output. However, such a statement, generic as it is, highly depends on a vast number of factors and problem parameters, such as the amount and quality of data available to us, the inherent hardness of the predictive task at hand, and the quality of the supervision, just to name a few. As a result, this promise is not always realized, and in practice successfully training a model that reaches that ideal requires copious amounts of experimentation with different hyperparameters and design choices. Thus, an emerging area in machine learning research is the one of theory-guided, or {\it physics-guided}, modeling \cite{karpatne2017theory}. It aims to leverage well-understood information from a particular scientific domain of interest, incorporating that in the design or the training of the model such that models which are consistent with what theory dictates are preferred over models that are inconsistent (with a varying degree of strictness, depending on how such guidance is incorporated). Doing so may imbue stability and efficiency in how the model is able to perform and generalize, thus making such guided approaches very appealing. In our scenarios, one may envision identifying parts of the instrument whose operation is well-understood and potentially expressed by precise equations, and enforcing consistency of the trained model with respect to those equations. The intended by-product of this action is that the model learned will capture phenomena (both documented and unknown) in the data with higher fidelity, even though full a-priori modeling of all phenomena may be infeasible or impossible.

\paragraph{Human (expert)-in-the-loop \& active learning}
\label{ml-active}
The typical view of supervision in machine learning considers the process of obtaining human annotations as an offline, slow, and tedious process. However, an emerging trend in machine learning is to blur the boundaries between those two processes by introducing ``human-in-the-loop’’ frameworks, where a model is being continually refined according to human feedback. Key challenges in doing so include the determination of when and how often to solicit feedback, the form of the feedback itself (e.g., it may be easier for a human labeler to identify whether or not two data points are of the same type than to provide a type for a single data point \cite{korlakai2016crowdsourced}), and in what ways the model should adapt to the new feedback provided by the human. Human-in-the-loop analytics and learning, albeit an active and emerging field, is related closely with ideas previously developed within the field of information retrieval (studying and developing algorithms for efficient search and retrieval of information) and the area of relevance feedback \cite{salton1990improving}. Furthermore, parts of the desiderata in a human-in-the-loop framework are also the subject of the area of active learning \cite{settles2009active}, where the objective is to solicit new annotations for a limited number of data points such that the new model’s performance is maximized. Active learning has found applications to science problems, such as anomaly detection in astronomy \cite{ishida2019active}. In realizing an end-to-end human-in-the-loop framework, however, more advances are necessary in a number of areas including scalability and visualization/human-computer interaction \cite{endert2014human,zhang2018name,xin2018accelerating}.
\section{Optimizing Instrument Operational Performance}
\label{instPerf}
Optimal instrument performance is critical for achieving the scientific goals of a large-scale experiment. However, the complexity of large instruments often makes it difficult to identify the {\it root cause} of errors encountered during operation. Normal operation typically involves controlling many coupled degrees of freedom simultaneously, with a multitude of potential points of failure and entry for noise. Using instrumental data to identify poor operating conditions and diagnose errors thus requires monitoring many interacting subsystems. Even so, modeling of the instrument, or even individual subsystems, often fails to capture enough realistic detail to reproduce operational errors because (1) the number of possible failure modes is vast and unknown and (2) the state of the instrument and all of its subcomponents, at any moment, is also not fully known. Thus, the best prospect for uncovering root causes of anomalous operation or failures lies in mining time series from the large number of diagnostic channels\footnote{A set of channels used for performing instrument diagnostics, primarily consisting of sensor readout or other quantities related to instrument control. \label{channeldef}}, various subsets of which may be non-trivially correlated, for \textit{interesting} patterns.

Among patterns of interest are the so called anomalies, they are outliers in the data that potentially hint towards processes that might be disrupting nominal operation of the instrument. In \cite{edelen2018opportunities}, the authors present a use-case of modern particle accelerators, a poster child for a large-scale complex instrument and cite that various anomaly detection techniques have been used at the LHC for identifying bad readings from beam position monitors and to assist in automated collimator alignment\cite{valentino2017anomaly}. A key challenge for human operators is selecting a small subset of relevant channels for investigation from a vast number of channels which record the dynamics of the instrument. In this section, we broadly describe two machine learning pipelines to aid the diagnosis of two types of operational issues - 1. transient noise and 2. control failures. We will first briefly introduce these operational issues and then present detailed real-world examples of these issues encountered in a \textit{state-of-the-art} complex instrument: ground-based gravitational wave detectors.

\paragraph{Transient noise} can contaminate the main data product of an experiment and thus lower its quality. For an active instrument like a particle accelerator this means repeating the experimental run which is costly but possible. For a passive observatory, like the ground-based gravitational-wave detectors at LIGO, such transient noise can lead to missing part or whole of a unique astrophysical event that is impossible to observe again. Figure \ref{fig:glitch} shows the now-famous example of a loud noise transient corrupting a portion of the signal from the first ever binary neutron start merger detected by LIGO. Thus, it is critical to understand sources of such noise to potentially eliminate it with upgrades. Using archived data and machine learning methods, we can identify \textit{witnesses} to noise artifacts by looking for correlations between transients in a set of diagnostic channels\footref{channeldef}. A subset of these diagnostic channels termed as \textit{witness channels} can then be used to categorize, locate and potentially eliminate the \textit{root cause} of a certain noise transients.

\paragraph{Control failures} render the instrument unable to operate in a nominal state and thus reduces duty-cycle. This operational issue has a relatively larger impact than the \textbf{transient noise} issue as no science data can be produced while the instrument is recovering. Diagnosing causes of control failures is crucial to mitigate future adverse events. Such failures are relatively easy to mitigate as they are occurring in real-time but a key task is to \textit{predict} an impending failure. We can address this task with machine learning by using data preceding failure events from a set of diagnostic channels to identify \textit{precursors} to these failure events. In \cite{felsberger2020explainable}, the authors explore the use of a deep neural networks \footnote{A class of supervised machine learning models that are particularly good at learning complex, non-linear functions of the input that map it to the prediction target.} for predicting operational failures at a particle accelerator facility by modeling precursors to a failure from a set of diagnostic channels\footref{channeldef}. \underline{\textit{Limited Supervision [\ref{ml-limsup}]:}} The machine learning problem is formulated as a binary classification task where data-points corresponding to failures (positive class) and nominal behaviour (negative class) can be automatically labelled without manual human effort. \underline{\textit{Explainable Models [\ref{ml-exp}]:}} Since deep neural networks are often black-box models, the authors employ an explanation technique, Layer-wise Relevance Propagation \cite{bach2015pixel}, to highlight a subset of diagnostic channels\footref{channeldef} that are relevant to failures.

\begin{figure}[t]
    \centering
    \includegraphics[width=1\textwidth]{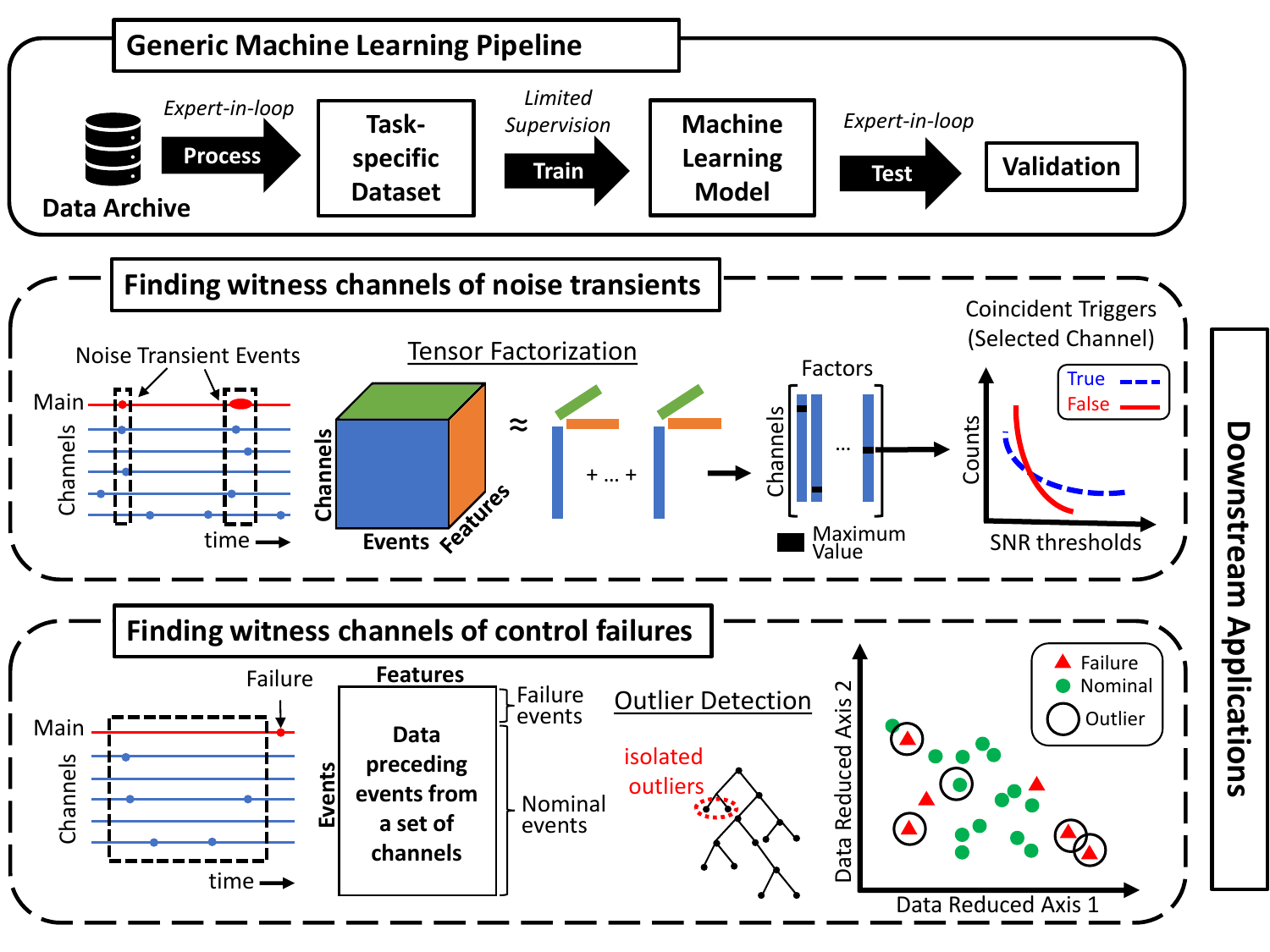}
    \caption{\textbf{\textit{Top row:}} A generic Machine Learning pipeline typically has 4 stages starting at task-specific dataset creation to validation of the modeling output. \textbf{\textit{Middle row:}} End-to-end pipeline for finding witnesses to noise transients that can potentially help diagnose sources of noise relies on selecting transient events of interest from the \textit{main} channel and finding coincidences in a set of diagnostic channels. \textbf{\textit{Bottom row:}} End-to-end pipeline for finding precursors of control failure events relies on isolating data from a set of witness channels preceding such events as outliers.}
    \label{fig:mlpipeline}
\end{figure}

\hfill \break
Before presenting real-world examples of the two aforementioned operational issues in context of our complex instrument of choice, LIGO, let us walk through the generic machine learning pipeline (shown in Figure \ref{fig:mlpipeline} (\textit{top row})). Raw data from a data archive is processed with a domain expert's guidance to create a \underline{\textbf{\textit{task-specific dataset}}}. This task-specific dataset is then used to train an appropriate \underline{\textbf{\textit{machine learning model}}}. The model's performance is tested on a holdout set of data-points that are not part the model's training for \textit{generalization} ability - a key expectation from any machine learning model. Finally, the \underline{\textbf{\textit{validation}}} of the model output is done using domain expert-defined tests which are often necessary for the downstream application of the model.

\subsection{Effect of noise transients on gravitational-wave searches}
\begin{figure}[t]
    \centering
    \includegraphics[width=0.70\textwidth,trim={0 4mm 7mm 5mm},clip]{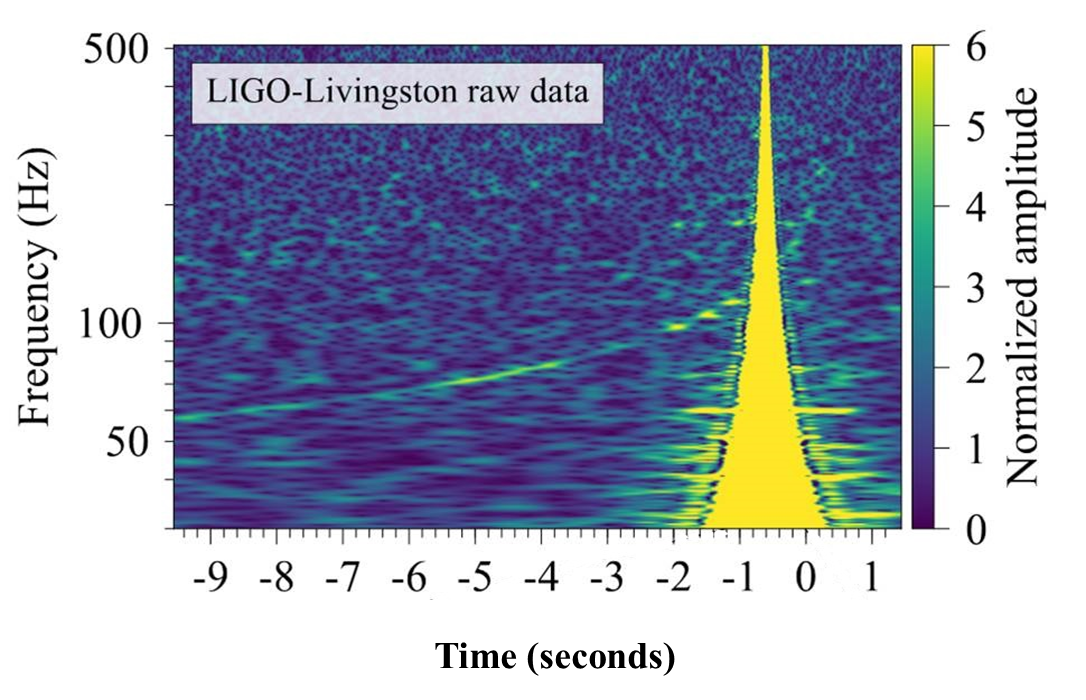}
    \caption{Time-frequency representation of the gravitational-wave event GW170817, as observed by the LIGO Livingston detector. Moments before merger, a noise glitch 1,000 times louder than the gravitational-wave signal occurred. Image reproduced courtesy of LIGO/Caltech/MIT.}
    \label{fig:glitch}
\end{figure}

For LIGO's online astrophysical search pipelines, one class of nonlinear noise artifact is particularly problematic: Transient noise bursts, of largely unknown origin, known as {\em noise glitches} \cite{Blackburn:2008} \cite{Cabero:2019}. Multiple glitches occur during a typical hour of observation. Glitches contaminate the astrophysical data streams (see, for example, Fig.~\ref{fig:glitch}), confusing burst-source gravitational-wave searches and hindering the timely issuance of real-time alerts for electromagnetic follow-up. By introducing a long non-Gaussian tail in the instrumental noise background, glitches also raise the statistical threshold for detecting true astrophysical events. Reducing the frequency of glitches will improve detection rates for all types of events, but most especially for high-mass binary black hole mergers, whose signal-to-noise ratio is the poorest.

\begin{figure}[t]
    \centering
    \includegraphics[width=\textwidth]{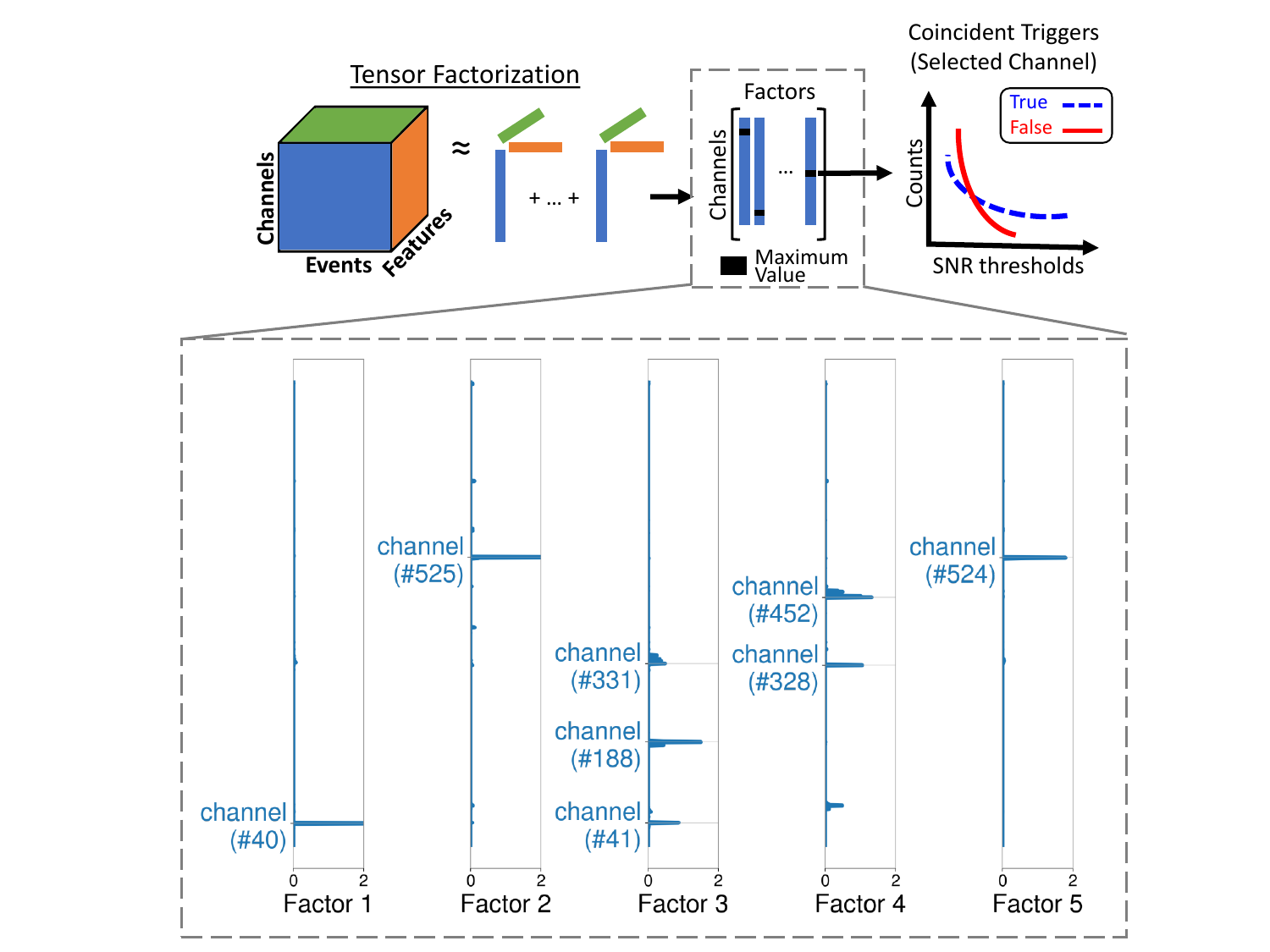}
    \caption{\textbf{\textit{Finding witness channels to noise transients: Witness Selection}} Factors corresponding to diagnostic channels of a $n=5$ component tensor factorization of $channels \times events \times features$ tensor that encodes presence or absence of coincident noise transients in diagnostic channels. They show high magnitude values in a factor for a few out of the approx. 900 channels used for this analysis. For example, Factor 1 shows a high magnitude value for channel \#40 which is then selected as a witness.}
    \label{fig:glitchfactors}
\end{figure}

\begin{figure}[t]
    \centering
    \includegraphics[width=\textwidth]{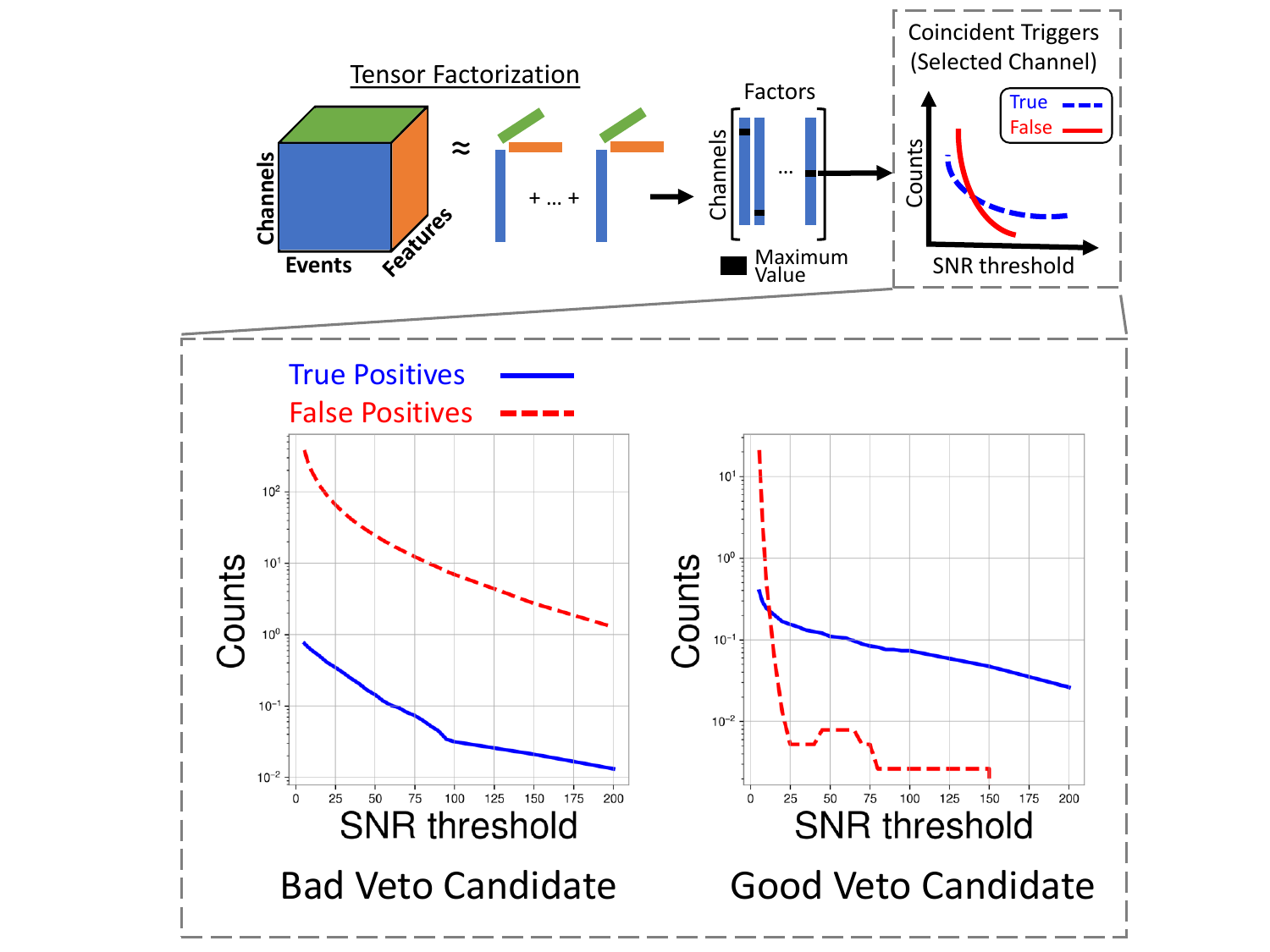}
    \caption{\textbf{\textit{Finding witness channels of noise transients: Witness Validation}} Two different channels selected using factors of the $channels \times events \times features$ tensor show differing true vs false positive characteristics. \underline{\textit{Left:}} The selected channel has a high false positive rate at increasing SNR thresholds compared to the true positive rate suggesting that it might just be an inherently noisier diagnostic channel and thus cannot be used to veto science data segments for downstream gravitational-wave search pipelines as it will result in large loss of science data. \underline{\textit{Right:}} There is a sharp drop-off in false positive rate of the selected channel above SNR threshold of 25 while the true positive rate remains relatively stable. We can use this selected channel to veto coincident noise transients in the science data.}
    \label{fig:fprtprvsSNR}
\end{figure}

In \cite{gurav2020unsupervised}, we use matrix and tensor factorization \cite{kolda2009tensor} to automatically identify relevant subsets from the set of diagnostic channels that are believed to be potential witnesses to \textit{glitches} present in LIGO data in a 5-day period from the third observing run. Figure \ref{fig:mlpipeline} (\textit{middle row}) shows an end-to-end pipeline for this task. We select a set of diagnostic channels and a set of \textit{glitches} from LIGO's main channel where gravitational-waves are observed. We then construct the \underline{\textbf{\textit{task-specific dataset}}} in form of a 3-mode tensor \footnote{In the context of machine learning, a tensor is a multidimensional array.} where mode-1 corresponds to the \textit{glitches} in the main channel, mode-2 corresponds to diagnostic channels selected for this analysis and mode-3 corresponds to features (e.g. duration, peak frequency, bandwidth, signal-to-noise ratio, etc.) of the \textit{glitches} found in the diagnostic channels that are coincident with the main channel \textit{glitches} within a short window around a main channel \textit{glitch}. This tensor essentially encodes the presence or absence of a \textit{glitch} in a diagnostic channel \textit{``coincident''} with a \textit{glitch} in the main channel. The \underline{\textbf{\textit{machine learning model}}} of choice is tensor factorization. We factorize this tensor into $N$ factors and obtain factor matrices corresponding to the latent space representations of each mode. We use the factor matrix corresponding to the ``diagnostic channels'' mode to select $N$ channels (one per factor), see Figure \ref{fig:glitchfactors}. We \underline{\textbf{\textit{validate}}} the channels thus selected as potential witnesses to the main channel \textit{glitches} in the dataset by examining the true positive rate\footnote{Number of times there was a coincident \textit{glitch} in a selected witness channel and the main channel divided by the number of total \textit{glitches} in the witness channel} and false positive rate\footnote{Number of times there was a \textit{glitch} in a selected witness channel but no coincident \textit{glitch} in the main channel divided by the number of total \textit{glitches} in the witness channel} of each selected witness channel for increasing values of signal-to-noise (SNR) thresholds, see Figure \ref{fig:fprtprvsSNR}. The channels deemed \textit{good veto candidates} by this validation step can then be considered for downstream applications like, for example, using the channel to veto and remove data segments from the main channel before searching for gravitational-wave signals.

\subsection{Effect of control failures on gravitational-wave detector duty cycle}

Achieving resonance in the full LIGO system is a complex process (see, e.g.,~\cite{Staley:2014}) typically requiring 30~minutes to complete. Control failure occurs when a disturbance, either internal or external, causes the laser cavities to become ``unlocked'' from their resonant operating points. This type of control failure is termed as \textit{lock loss}. Due to the frequency of lock losses, all three LIGO-Virgo detectors (LIGO Livingston, LIGO Hanford, and Virgo) were running at the same time only about 50\% of the time during the third observing run. The majority of lock losses have no readily identifiable environmental cause~\cite{Rollins:2017}, so may too be triggered by non-Gaussian noise events. Identifying and eliminating triggers of lock loss could potentially make a big improvement in the fraction of calendar time all three detectors are data taking, increasing the number of triple-coincident gravitational-wave detections. Observations by multiple detectors with different antenna patterns are critical for precise sky localization, as needed for targeted electromagnetic follow-up of potential multi-messenger events.

\begin{figure}[t]
    \centering
    \includegraphics[width=\textwidth]{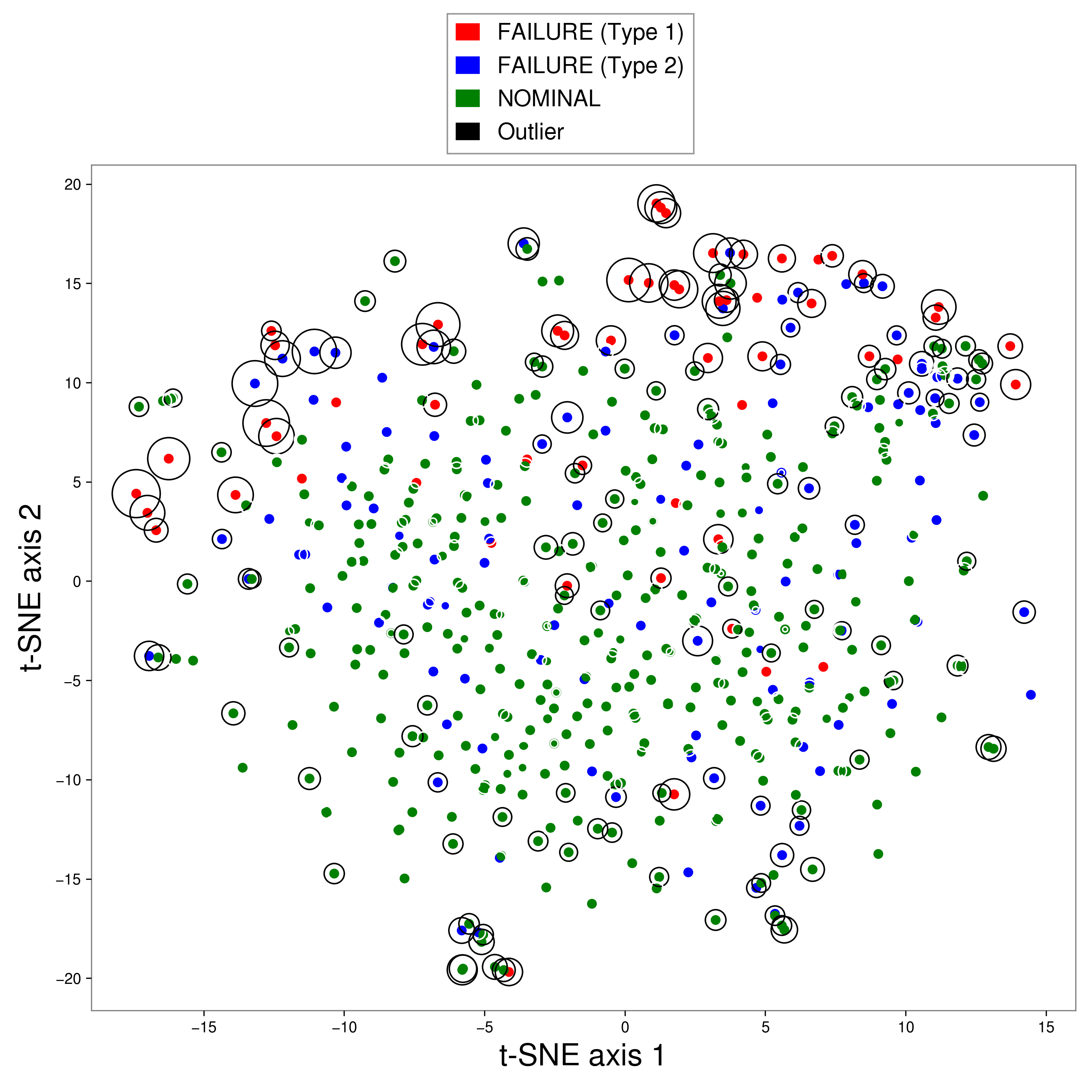}
    \caption{\textbf{\textit{Finding witness channels of control failures:}} (Note: The figure above shows a 2-dimensional projection of the $m$-dimensional data ($m >> 2$) using t-Distributed Stochastic Neighbor Embedding (t-SNE) \cite{JMLR:v9:vandermaaten08a} that tries to preserve distances in the higher dimensional space for the purpose of visualization only). We obtain $k$ features from segments of data preceding 2 types of control failures (\textit{red and blue}) from a set of $n$ diagnostic channels ($m=k \times n$). To complete the dataset we also sample data segments during nominal operation (\textit{green}). The outlier detection algorithm is fully unsupervised and does not use the ground-truth labels for the events in the dataset (failure or nominal) to isolate certain events as outliers (\textit{circled in black}). We can see that instances of data preceding control failures (\textit{red, blue}) are identified as outliers. The diameter of the black circle in this plot corresponds to an outlier score (larger score $=$ bigger outlier) and this score can potentially be used as a threshold for reducing false positives (nominal events that are identified as outliers but with a relatively smaller outlier score).}
    \label{fig:locklossoutlier}
\end{figure}

Figure \ref{fig:mlpipeline} (\textit{bottom row}) shows a pipeline that formulates the problem of diagnosing control failures as an anomaly detection task where the \underline{\textbf{\textit{task-specific dataset}}} is constructed by obtaining data preceding control failures from a set of diagnostic channels. We also obtain data during nominal operation periods from these channels. Thus, we do not need any explicit \textit{supervision} as we have automatic \textit{ground-truth (labels)} for the data-points in our dataset as either \textit{failure} or \textit{nominal} events. We hypothesize that data preceding failure events in a subset of channels will have anomalies which might be precursors to the failures. The \underline{\textbf{\textit{machine learning model}}} of choice here is an unsupervised, tree-based Outlier Detection algorithm called Isolation Forest \cite{liu2008isolation} that partitions the dataset using a tree structure and \textit{isolates} anomalies as outliers. We validated the events deemed as anomalies by the algorithm by comparing them against the ground-truth, see Figure \ref{fig:locklossoutlier}.

\hfill \break
There are some key challenges worth highlighting in the machine learning pipelines described in Section 1.3.1 and 1.3.2. Firstly, we assume a preprocessing step of creating the task-specific datasets from the massive amounts of raw, archived time series data. This step has an inevitable need of domain expertise as there can be potentially tens of thousands of diagnostic channels to choose from only a subset of which may be relevant to the task and these channels are often sampled at different rates capturing various phenomenon occurring at different time and frequency scales. Beyond the selection of a set of diagnostic channels for any given analysis, the \textit{``features''} engineered from the raw data-streams can be explicitly hand-crafted to be semantically meaningful to the domain-expert or they can be implicitly learned by the machine learning model. Implicit feature representations of our data, like the ones learned by black-box models like deep neural networks, reduce the burden of hand-crafting good enough features but make the output of our pipelines less interpretable. Moreover, at the end of the pipeline we also seek expert guidance for creating validation tests that examine the utility of the machine learning model's output for downstream diagnostic applications.

\subsection{Public Datasets}

Most real-world complex instruments have large quantities of raw archival operational data, as described in this section, that is not readily available to be used for training machine learning models. Therefore, machine learning applications use smaller benchmark datasets curated from the data archive or create synthetic datasets for \textit{proof-of-concept}.

Following are some LIGO-specific public datasets available for machine learning in complex instruments diagnostics.
\begin{enumerate}
    \item \textbf{LIGO Gravity Spy:} This dataset, as described in \cite{zevin2017gravity}, consists of time-frequency spectrogram images of various noise transients in the main data product of LIGO. The machine learning task is to classify these images into a set of classes using state-of-the-art deep learning models like convolutional neural networks. The sample dataset is available here - \url{https://zenodo.org/record/1476551#.YjoQ-3XMKV5}
    
    \item \textbf{LIGO diagnostic channels dataset:} This is a multivariate timeseries dataset consisting of a 3-hour period centered on a real gravitational-wave event, GW170814, from approx. 1,000 LIGO diagnostic channels. This dataset can be used for identifying noise transients witness channels as described in this section or for subtracting noise from the main data product. The dataset can be found at the Gravitational Wave Open Science Center webpage - \url{https://www.gw-openscience.org/auxiliary/GW170814/}
\end{enumerate}
\section{Optimizing Instrument Design}
\label{instDesign}

A fundamental requirement in instrument design is the ability to accurately and efficiently simulate different scenarios for a variety of instrument configurations, in pursuit of configurations which optimize a certain target physical quantity. In this section, we first provide a brief overview of the traditional simulation and optimization pipeline, and subsequently we review an emerging machine learning breakthrough which has the potential to revolutionize instrument design.

\subsection{Current simulation \& optimization tools}

Traditionally, the instrument design pipeline is roughly divided into two parts, simulation, where the behavior of an instrument (or one of its components) is estimated by software, and optimization, where the designer seeks to identify the best combination of parameters that satisfies a set of requirements for the component in question.

\paragraph{Simulation}
Simulation is the process of estimating the response of an instrument or one of its components through software, thus eliminating the need to fabricate and physically experiment with that component. Such simulators have been widely used by both accelerators \cite{agostinelli2003geant4} and gravitational wave detectors \cite{freise2013finesse}. Typically, during simulation, the user/designer specifies a thorough description of the component that is being simulated, and provides as input the environmental conditions (e.g., a certain event of interest) for which the component’s response is needed. In turn, the simulator computes the response of that component, usually by numerically solving for the equations which govern that component’s response. Finally, the last step in the pipeline takes as input the response/state of the component, and outputs a reconstruction of a quantity of interest, which is used as an evaluation metric (e.g., energy).

\paragraph{Optimization}
During a single run of the simulator, the parameters of the design to be simulated are user-defined. However, manually iterating over different designs in pursuit for one or more designs which satisfy the chosen criteria can be extremely inefficient and incomplete. To that end, in addition to simulators, effective design uses optimizers which are given a user-specified objective to optimize for (e.g., energy) and their goal is to effectively navigate the search space of designs and identify one or more that achieve the optimal objective. Ocelot \cite{tomin2017line} is such a general-purpose optimization tool which has been very popular in accelerator design. Interestingly, there are connections between optimization tools in instrument design and tools used in modern machine learning to fine-tune the design parameters (also called hyperparameters) of a large model, such as Bayesian Optimization \cite{mcintire2016bayesian,turner2021bayesian}.

\subsection{Generative models for instrument simulation}
Despite the existence of efficient optimizers, whose goal is to minimize the number of times the simulation must be run, simulation in itself is a very computationally intensive and slow process. Furthermore, in order for a simulation to be accurate, it is typically restricted to very specific components of a larger instrument. As a result, simulating an instrument end-to-end would require the combination of a large number of specialized simulators, which are not necessarily designed to be compatible with each other. Thus, such an endeavor may be technically challenging or even intractable.

On the other hand, machine learning models, and especially deep neural networks, can be used as cheap, generic nonlinear function approximators. They have been studied for ``data analysis'' of an instrument's main data products (e.g., using deep neural networks to quickly estimate astrophysical parameters of a gravitational wave signal in real time, instead of carrying out the exact numerical calculations \cite{george2018deep}). However, an emerging trend in instrument design is to replace traditional simulators with deep learning models, specifically leveraging recent advances in generative adversarial networks (GANs) \cite{goodfellow2020generative}.

A GAN is a machine learning model that itself consists of two sub-models, the {\it generator} and the {\it discriminator}. Typically those two models are neural network based, so the term ``network'' is commonly used when referring to them. The generator takes as input a vector, typically drawn at random from a Gaussian distribution, and it is tasked with outputting a data point (e.g., an image or the output of an instrument simulator) that is ``realistic.’’ The discriminator, on the other hand, takes two inputs: a ``real’’ data point (e.g., an actual image in our dataset or an actual output of an instrument simulator) and a generated data point (the output of the generator), and its task is to determine which data point is real and which is fake. During training, those two networks are engaged in a two-player game, where, as the generator gets better at producing realistic outputs, the discriminator improves its ability to detect fake data, and vice-versa. At the end of the training process, the generator will ideally produce data indistinguishable from ``real’’ data, while effectively providing a means of sampling from the distribution of ``real’’ data.

\begin{figure}[!ht]
    \centering
    \includegraphics[width = 0.99\textwidth]{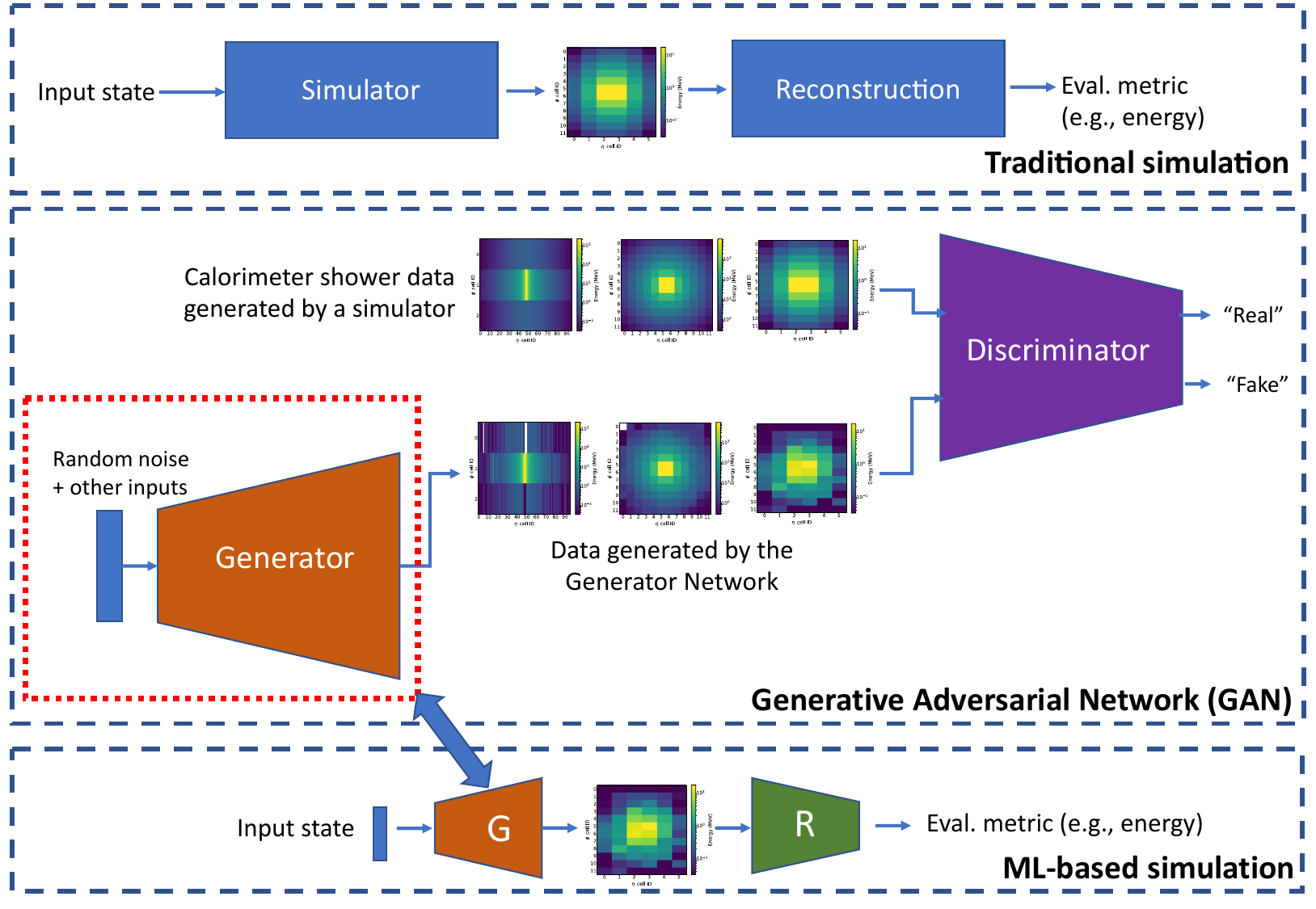}
    \caption{{\em Top row}: An illustration of the ``traditional'' simulation pipeline\\
    {\em Middle row}: Illustration of a Generative Adversarial Network (GAN): The Generator Network is taking as input random noise and additional information about the state to be simulated and generates ``fake'' data. The Discriminator Network takes as input ``real'' data (as simulated by a traditional simulator such as geant4) and ``fake'' data generated by the Generator, and determines which data point is fake or real.\\
    {\em Bottom row}: In the ML-based simulation pipeline, the simulator is replaced by the GAN Generator and a ``Reconstruction'' neural network is replacing the analytical reconstruction process, which transforms the simulation output to a quantity of interest (e.g., energy). (Data figures show simulated and generated calorimeter showers and are adapted from \cite{paganini2018accelerating}).}
    \label{fig:GAN_fig}
\end{figure}

The most well-known use case and success story of GANs is in the generation of images and videos that look extremely realistic \cite{karras2019style}. However, the success of GANs extends far beyond this use case, with successful applications in drug discovery \cite{maziarka2020mol} and material design \cite{dan2020generative}. Very recently, there have been successful attempts at using GANs for the simulation of complex scientific instruments, as applied  specifically to the simulation of accelerator calorimeters \cite{de2017learning,paganini2018accelerating,note2020fast}. Figure \ref{fig:GAN_fig} provides an illustrative example. These recent attempts clearly demonstrate the feasibility and potential of harnessing the power of GANs for complex instrument design.

Despite early success on the calorimeter application, broad application of GANs to the problem of instrument simulation and design poses fascinating interdisciplinary research challenges. Those challenges pertain to ensuring that the generated data are of high quality, both as it pertains to how realistic they are as well as to the diversity of generated data points (a known mode of failure for GANs is memorizations of the training set or generation of very similar data points) \cite{shmelkov2018good}. Advances in the following areas, outlined in Sec. \ref{MLConcepts}, can pave the way for tackling those challenges:

(a) {\em Physics-guided models}: In order to ensure that the Generator will obey physical constraints and laws governing the instrument that it models, a physics-guided approach can incorporate this information in the design or training stages of the model. In Paganini et al. \cite{paganini2018accelerating}, the proposed GAN takes such a physics-guided approach, by modifying the loss function that is being optimized during training.

(b) {\em Limited supervision}: Training data for the GAN come from very expensive simulations. Thus, it may be infeasible to create very large training data sets, a challenge which may amplify issues such as the memorization of the training set by the GAN. Thus, designing solutions that work with limited supervision is imperative.

(c) {\em Expert-in-the-loop}: A big advantage of applying GANs to instrument simulation is that the definition of ``realistic’’ is objective and can be analytically measured. This is in contrast to the original application of GANs, where judging the realism of a generated face or natural image can be subjective. However, when judging the diversity of generated data, the involvement of human experts can be crucial.

In addition to the immediate efficiency gains of substituting an inefficient simulator with a highly-efficient neural network, a major promise of machine learning driven instrument design is flexibility and modularity: once every component of an instrument can be efficiently simulated by a neural network, this can facilitate the simulation of increasingly larger portions of an instrument, ideally all the way to end-to-end simulation and design.

\subsection{Public Datasets}

Three datasets used in \cite{atlas2021atlfast3} for simulating particle showers using GANs at the ATLAS detector at the LHC are hosted here - \url{https://opendata.cern.ch/record/15012}
\section{Conclusion}
\label{conc}
Instruments at the frontiers of science are becoming increasingly complex and data-intensive. There is a need for intelligent automation to tame this complexity, increase reliability and push the limits of these technological marvels to make new scientific discoveries. This chapter explored the emerging application of machine learning for complex instrument operations management and design. The current state of the art demonstrates very promising results, with machine learning methods empowering early detection of failures, diagnosing noise sources, and enabling the flexible simulation of different components of an instrument---tasks which may have previously been extremely tedious or downright impossible to do well by hand. There exist a number of fundamental challenges within machine learning research to successfully apply these techniques to solving issues of complex instruments, like the need for explainable and theory-guided modeling. Successful application of these techniques can pave the way for a new generation of complex instruments.







\bibliographystyle{ws-book-van}    
\bibliography{references}      




\end{document}